# Quasi-analytical solutions of wide-angle and broadband insulator-metal grating-metal metamaterial absorber in visible to near-infrared bands

PIYAWATH TAPSANIT*, CHESTA RUTTANAPUN

*Department of Physics, Faculty of Science, King Mongkut's Institute of Technology Ladkrabang, Ladkrabang, Bangkok, 10520, Thailand*
*Corresponding author: p_epsilon@yahoo.com*

**A metamaterial absorber converts solar radiation into heat for applications in thermoelectric generators and solar thermophotovoltaics. A metamaterial absorber typically consists of metallic parts separated by an insulator layer which does not allow the total heat extraction via direct heat conduction. Here, we propose wide-angle and broadband metamaterial absorber comprising an insulator layer, metal grating, and metal back plane (IMM) that allows the total heat extraction via direct heat conduction. We have formulated the quasi-analytical solution (QANS) of the IMM-absorber by applying the coupled-mode analysis with new technique for taking finite dielectric constant of a metal into account. The QANS are consistent with finite difference time domain (FDTD) simulation. We fine that the IMM-absorbers made by tungsten and silver metals give the best and worse average absorbance over the wavelength range 400-1,200 nm, respectively. The best IMM-absorber shows the optimized average absorbance 96.2%. The broadband absorption results from the excitations of insulator-metal waveguide resonance and slit waveguide resonance that are insensitive to incident angles. The averaged absorbance is above 90.0% for the incident angles 0-66 degree. The IMM-absorber performance may be enhanced by using metal rod array with engineered supperlattice.**

http://dx.doi.org/10.1364/AO.99.099999

Solar energy is free and abundant energy source. A solar absorber converts the solar energy into heat which is important in thermoelectric power generators and in solar thermophotovoltaics [1,2]. The generated heat could be stored by phase-change materials for uses at night [3]. Good solar absorber must absorb radiations over broad optical spectrum, and it must be independent on incident angles and polarizations. The absorber made by vertically aligned cabonnanotubes (Vatablack) is known as the best absorber due to its high absorption over 99% and the independence on incident angles over UV to infrared bands [4]. However, the vantablack is relatively expensive which is not suitable for large-scale applications [5]. By using cheaper materials, the large-scale absorber can be affordable. Instead of inventing new materials, naturally available materials can be engineered into subwavelength periodic structures to attain the absorber requirements. These engineered devices are called metamaterial absorbers [6]. The broadband metamaterial absorber typically comprises metal nanostructure and metal back plane separated by an insulator layer [7,8]. The broadband absorption arises from the angle-insensitive magnetic resonances resulting from the complete circulation of electric current [9]. Recently, iron nanostructure has been applied to this type of absorber [10]. The iron absorber shows the averaged absorbance 97% over the wavelength range 400-1,500 nm [10]. However, heat is generated in both metal nanostructure and metal back plane [8], and thus heat in the metal nanostructure cannot directly conduct to the metal back plane hindering the efficient extraction of the total heat.

Here, we propose wide-angle, broadband, and nearly-perfect absorber comprising lossless insulator, metal grating, and metal back plane which is called IMM-absorber. The metal grating is in contact with the metal back plane allowing the extraction of total heat via heat conduction. We have formulated new quasi-analytical solution (QANS) of the IMM-absorber by modifying the coupled-mode analysis [11] taking into account the finite dielectric constant of the metal by using the metallic admittance. The QANS are confirmed by finite difference time domain (FDTD) simulation using freely-distributed program called MEEP [12]. Then, the QANS are used to optimize the IMM-absorber by varying its geometrical parameters and metals. We also explain the physical origins of the broadband absorption and discuss the ways to improve the IMM-absorber performance.

## 1. Calculation method

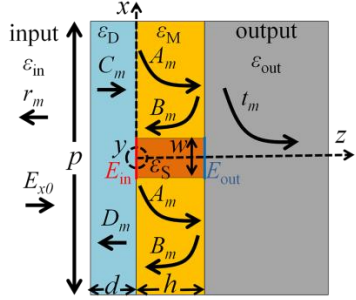

Fig. 1. Schematic drawing of the IMM-absorber's unit cell comprising an insulator layer (blue color) placed on metal grating (gold color) which is backed by metal output plane (grey color). The dielectric constants of all regions, geometrical parameters, and field coefficients are indicated in the figure, and their meanings are described in the main text.

The schematic drawing of the IMM-absorber's unit cell over the $xz$-plane is shown in Fig. 1. The absorber is periodic along $x$-axis and homogeneous along $y$-axis. The metallic part of the metal grating has dielectric constant $\varepsilon_M$. The slit has height $h$ and width $w$. The period of the slit array along $x$-axis is denoted by $p$ and each slit may be filled with medium with dielectric constant $\varepsilon_S$. The output medium defined as a metal with dielectric constant $\varepsilon_{out}$ is placed in contact with the metal grating's output side. Thickness of the output back plane could be a few hundred nanometers so that the visible and infrared lights can be completely blocked. An insulator layer with dielectric constant $\varepsilon_D$ and thickness $d$ is placed in contact with the metal grating's input side. The incident light is propagating through the input medium with dielectric constant $\varepsilon_{in}$. In this work, we will fix $p=300$ nm and $h=100$ nm, but $d$ and $w$ will be optimized by varying their values from 1 nm to 200 nm for $d$ and from 1 nm to 180 nm for $w$. An alumina with $\varepsilon_D=2.89$ is used as the lossless insulator because of its high melting point. An input medium and slit waveguide are defined as air. The metallic part of metal grating and the output medium are the same metallic materials. We consider Ag, Al, Au, Cr, Cu, Ni, Pd, Pt, Ti, and W metals whose dielectric constants are given by A. D. Rakic et al [13].

The IMM-absorber is excited by transverse-magnetic (TM) wave having magnetic field polarized along $y$-axis and electric field polarized over $xz$-plane. The time harmonic exp($-i\omega t$) is assumed. The incident $E_x$ has amplitude $E_{x0}$, and the incident $E_z$ is determined by the Gauss's law. The incident $H_y$ is determined by the incident electric field by using the Faraday's law.

The $m$th-order reflection (transmission) coefficient denoted by $r_m$ ($t_m$) is defined as fraction of $E_{x0}$ being reflected (transmitted) to the $m$th-order Bloch mode in the input (output) medium, where $m$ is an integer. Then, the reflected (transmitted) $E_x$ is generally written as the summation of $r_m\langle x|\beta_m\rangle$ ($t_m\langle x|\beta_m\rangle$), where $\langle x|\beta_m\rangle = e^{i\beta_m x}/\sqrt{p}$, $\beta_m = k_x + m2\pi/p$ is normalized $m$th-order Bloch mode. The remaining components of electric and magnetic fields are obtained in the same way as the TM incident wave. Similarly, $E_x$ inside an insulator is the summation of TM waves propagating out from the interface $z=-d$ with field coefficient $C_m$ and TM waves propagating out from the interface $z=0$ with field coefficient $D_m$. For the metal grating, there are two parts: (i) slit waveguide, and (ii) metallic part. In the slit waveguide, it is assumed that $w<<\lambda$ and electric field is zero at the slit waveguide's walls so that the slit's fundamental waveguide mode has constant $E_x$ along $x$-axis and zero $E_y$. Therefore, the normalized slit's fundamental waveguide mode can be defined as $\langle x|\alpha\rangle = 1/\sqrt{w}$. It is convenient for the coupled-mode analysis to define $E_{in}$ as the amplitude of $E_x$ at the slit's opening and $E_{out}$ as the negative amplitude of $E_x$ at the slit's exit. In the metallic part of the grating, $E_x$ is the summation of TM waves decaying out from the interface $z=0$ with field coefficient $A_m$ and TM waves decaying out from the interface $z=h$ with field coefficient $B_m$. It is assumed that $h$ is sufficiently larger than the metal's skin depth so that the former (latter) TM waves approximately vanish at $z=h$ ($z=0$). It should be noted that our technique is different from the previous method that applied the surface impedance boundary condition (SIBC) at $z=0$ and $z=h$ interfaces [11]. According to J. D. Jackson [14], the SIBC neglects $E_z$ inside the metal. However, our technique uses metallic admittance that takes both $E_x$ and $E_z$ inside the metal into account which may allow more accurate calculation of electric field amplitude inside the metal than the previous method.

All field coefficients are linear function of $E_{in}$ and $E_{out}$. The continuities of $E_x$ and $H_y$ at interfaces of the input and output sides of the metal grating, combined with the coupling processes, give the solutions of $E_{in}$ and $E_{out}$ written in terms of the input and output coupling parameters denoted by $G_{in}$ and $G_{out}$, respectively [11]. We will describe the processes used to obtain the linear equation of $E_{in}$ and $E_{out}$ of the input side of the metal grating as follows, while another linear equation of the output side of the metal grating can be obtained by applying the same steps.

Firstly, we consider the continuities of $E_x$ and $H_y$ at the interface $z=0$. The continuity of $H_y$ only on the metallic part of the metal grating, excluding the slit waveguide, gives $A_m$ written in terms of $C_m$ and $D_m$ as follows

$$A_m = \left(Y_m^{(D)}/Y_m^{(M)}\right)\left(C_m \exp\left(iq_m^{(D)}d\right) - D_m\right), \quad (1)$$

where $q_m^{(D)} = \sqrt{k_D^2 - \beta_m^2}$ is $m$th-order $z$-component of wavevector in an insulator layer, $k_D = \sqrt{\varepsilon_D}k_0$, $Y_m^{(D)} = \varepsilon_D k_0/q_m^{(D)}$ is $m$th-order admittance in an insulator layer, $Y_m^{(M)} = \varepsilon_M k_0/q_m^{(M)}$ is $m$th-order admittance in the metallic part of grating, $q_m^{(M)} = \sqrt{k_M^2 - \beta_m^2}$ is $m$th-order $z$-component of wavevector in the metallic part of grating, and $k_M = \sqrt{\varepsilon_M}k_0$. Next, the continuity of $E_x$ is imposed at the same interface including both metallic part and slit waveguide of the metal grating, and we obtain the following equation

$$A_m + E_{in}\langle\beta_m^{(x)}|\alpha\rangle = C_m \exp\left(iq_m^{(D)}d\right) + D_m, \quad (2)$$

where $\langle\beta_m^{(x)}|\alpha\rangle$ denotes the inner product between the $m$th-order Bloch wave and the slit waveguide mode. By substituting Eq. (1) into (2), we obtain the following equation

$$\rho_{M,D}^{(m)}C_m \exp\left(iq_m^{(D)}d\right) + D_m = (1/2)\tau_{M,D}^{(m)}E_{in}\langle\beta_m^{(x)}|\alpha\rangle, \quad (3)$$

where $\rho_{\alpha,\beta}^{(m)}$ and $\tau_{\alpha,\beta}^{(m)}$ denote, respectively, $m$th-order reflection and transmission coefficients from the media $\alpha$ to $\beta$ defined as follows [15]

$$\rho_{\alpha,\beta}^{(m)} = \frac{Y_\alpha^{(m)} - Y_\beta^{(m)}}{Y_\alpha^{(m)} + Y_\beta^{(m)}}, \quad \tau_{\alpha,\beta}^{(m)} = \frac{2Y_\alpha^{(m)}}{Y_\alpha^{(m)} + Y_\beta^{(m)}}. \quad (4)$$

Next, the continuity of $H_y$ at the same interface $z=0$ is imposed again but now only on the slit waveguide, and we obtain the following equation

$$-iY_m^{(D)} C_m \exp(iq_m^{(D)} d) + iY_m^{(D)} D_m = (\gamma E_{in} + G_V E_{out})\langle \beta_m^{(x)} | \alpha \rangle, \quad (5)$$

where $\gamma = (k_S / k_0)\cot(k_S h)$ and $G_V = (k_S / k_0)\csc(k_S h)$ denote, respectively, the multiple reflections of TM waves inside the slit waveguide and the coupling of TM waves on two faces of the metal grating via the slit waveguide [16], and $k_S = \sqrt{\varepsilon_S} k_0$. Equations (4) and (5) can be wrapped in the matrix form as

$$\begin{bmatrix} \rho_{M,D}^{(m)} e^{iq_m^{(D)}d} & 1 \\ -iY_m^{(D)} e^{iq_m^{(D)}d} & iY_m^{(D)} \end{bmatrix} \begin{pmatrix} C_m \\ D_m \end{pmatrix} = \begin{bmatrix} \tau_{M,D}^{(m)} & 0 \\ \tau & G_V \end{bmatrix} \begin{pmatrix} E_{in} \\ E_{out} \end{pmatrix} \langle \beta_m^{(x)} | \alpha \rangle. \quad (6)$$

Next, $(C_m \; D_m)^T$ can be written in terms of $r_m$ and the incident light by imposing the continuities of $E_x$ and $H_y$ at the interface $z=-d$. This leads to the following matrix equation

$$\begin{pmatrix} C_m \\ D_m \end{pmatrix} = \frac{1}{\tau_{D,in}^{(m)} e^{iq_m^{(D)}d}} \begin{bmatrix} e^{iq_m^{(D)}d} & \rho_{D,in}^{(m)} e^{iq_m^{(D)}d} \\ \rho_{D,in}^{(m)} & 1 \end{bmatrix} \begin{pmatrix} \delta_{m0} e^{-ik_z^{(in)}d} \\ r_m \end{pmatrix}. \quad (7)$$

By substituting Eq. (7) into (6), we obtain new matrix equation whose first row gives the solution of $r_m$ written as follows

$$r_m = -\left( \frac{\rho_{D,in}^{(0)} + \rho_{M,D}^{(0)} e^{2i\phi_0^{(D)}}}{1 + \rho_{M,D}^{(0)} \rho_{D,in}^{(0)} e^{2i\phi_0^{(D)}}} \right) \delta_{m0} e^{-ik_z^{(in)}d} + \frac{1}{2} \frac{\tau_{M,D}^{(m)} \tau_{D,in}^{(m)} e^{i\phi_m^{(D)}} E_{in} s_m}{1 + \rho_{M,D}^{(m)} \rho_{D,in}^{(m)} e^{2i\phi_m^{(D)}}}, \quad (8)$$

where $s_m = \langle \beta_m^{(x)} | \alpha \rangle = \sqrt{w/p} \, \mathrm{sinc}(\beta_m w/2)$, and second row, with the coupling process, gives the linear equation of $E_{in}$ and $E_{out}$ written as follows

$$(G_{in} - \gamma) E_{in} - G_V E_{out} = I_0, \quad (9)$$

where $G_{in}$ denoting the input-coupling parameter is defined as follows

$$G_{in} = i\sum_m \frac{Y_m^{(D)} Y_m^{(M)}}{Y_m^{(D)} + Y_m^{(M)}} \left( \frac{1 - \rho_{D,in}^{(m)} e^{2i\phi_m^{(D)}}}{1 + \rho_{M,D}^{(m)} \rho_{D,in}^{(m)} e^{2i\phi_m^{(D)}}} \right) s_m^2, \quad (10)$$

where the summation is run over the increasing order $m$ until the value is converged, and $I_0$ denoting the direct excitation is given as follows

$$I_0 = \frac{iY_0^{(D)} \tau_{M,D}^{(0)} \tau_{in,D}^{(0)} e^{i\phi_0^{(D)}}}{1 + \rho_{M,D}^{(0)} \rho_{D,in}^{(0)} e^{2i\phi_0^{(D)}}} e^{-ik_z^{(in)}d} s_0. \quad (11)$$

Another linear equation of $E_{in}$ and $E_{out}$ and the solution of $t_m$ are obtained by applying the continuities of $E_x$ and $H_y$ at the interface $z=h$ by using the same steps as those of the input side. These processes give the solution of $t_m$ written as follows

$$t_m = -\tau_{M,out}^{(m)} E_{out} s_m / 2, \quad (12)$$

and the linear equation of $E_{in}$ and $E_{out}$ written as follows

$$-G_V E_{in} + (G_{out} - \gamma) E_{out} = 0, \quad (13)$$

where $G_{out}$ denoting the output-coupling parameter is defined as follows

$$G_{out} = i\sum_m \frac{Y_m^{(M)} Y_m^{(out)}}{Y_m^{(M)} + Y_m^{(out)}} s_m^2. \quad (14)$$

It should be noticed that $G_{in} = G_{out}$ when an insulator layer is remove and the input and output media are made from the same material.

By solving Eq. (9) and (13), we obtain the solutions of $E_{in}$ and $E_{out}$ written as follows

$$E_{in} = \frac{(G_{out} - \gamma) I_0}{(G_{in} - \gamma)(G_{out} - \gamma) - G_V^2}, \quad (15)$$

$$E_{out} = \frac{G_V I_0}{(G_{in} - \gamma)(G_{out} - \gamma) - G_V^2}. \quad (16)$$

The $0$th-order reflectance $R$ is obtained from $r_0$ by $R = 1 - |r_0|^2$, and the $0$th-order absorbance $A$ is obtained from $R$ by $A = 1 - R$ since the $0$th-order transmittance is zero.

Heat is generated only in the metallic parts of the IMM-absorber. The rate of generated heat density $Q_d$ can be obtained by $Q_d = (1/2) \mathrm{Re}(\sigma) |\mathbf{E}|^2$, where $\sigma$ is complex conductivity of a metal [17]. By using the relation $\mathrm{Re}(\sigma) = \varepsilon_0 \omega \mathrm{Im}(\varepsilon)$ [17], we obtain $Q_d = (1/2) \varepsilon_0 \omega \mathrm{Im}(\varepsilon) |\mathbf{E}|^2$ [8]. Therefore, $Q_d$ normalized by the incident power density $S_0$ ($z$-component of the Poynting vector) can be computed by using the following relation

$$Q_d / S_0 = k_0 \mathrm{Im}(\varepsilon_M) |\mathbf{E}/\mathbf{E}^{(inc)}|^2, \quad (17)$$

where $\mathbf{E}$ is the electric field in the metallic part of the IMM-absorber and $\mathbf{E}^{(inc)}$ is the incident electric field.

## 2. Results

### A. Optimized IMM-absorber and its spectrum

The IMM-absorber is optimized by QANS which are used to fine the maximum of the average absorbance over the wavelength range 400-1,200 nm for specified ranges of $w$ and $d$ as previously mentioned. The black dashed line in Fig. 2(a) shows the $0$th-order absorption spectrum of the optimized IMM-absorber made by tungsten (W) which is well consistent with that from FDTD as shown by the red solid line. The average absorbance is high and equal to 96.2% which is about 1% smaller than that of the MIM-absorber [10]. Two broad peaks appear at wavelengths 553 nm (2.242 eV) and 1,121 nm (1.106 eV). The relatively large difference between QANS and FDTD is noticed at long wavelength due to poor plasmonic response of W.

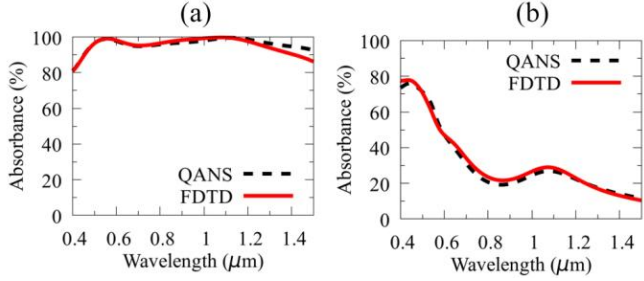

Fig. 2. Absorption spectra of IMM-absorber made by tungsten (a), and gold (b) with geometrical parameters $w$=71 nm and $d$=75 nm. The black dashed line is QANS spectrum and the red solid line is FDTD spectrum.

Better consistent result at long wavelength is expected by taking into account non-zero electric field at the slit waveguide's walls or by using a new metal with better plasmonic response.

Figure 2(b) shows the $0$th-order absorption spectra of the IMM-absorber made by gold (Au) with the same geometrical parameters as those used in Fig. 2(a). The more consistent result is obtained at long wavelength due to good plasmonic response of Au. However, Au has small loss at near-infrared band (high reflection at infrared band), and the average absorbance of Fig. 2(b) over the wavelength 400-1,200 nm is only 36.4%, about 2.6 times lower than W IMM-absorber.

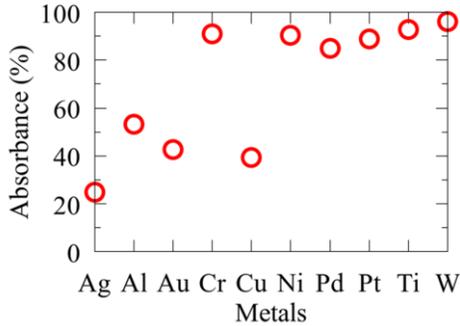

Fig. 3. Average absorbance of optimized IMM-absorbers made by various metals.

Figure 3 shows the averaged absorbance of the optimized IMM-absorbers made by various metals. The optimized average absorbance of the IMM-absorbers made by Ag, Al, Au, Cr, Cu, Ni, Pd, Pt, Ti, and W are equal to 24.9%, 53.3%, 42.7%, 91.1%, 39.5%, 90.5%, 85.1%, 88.9%, 92.8%, and 96.2%, respectively, and the corresponding optimized pair ($w,d$) for each IMM-absorber are equal to (12,81) nm, (25,89) nm, (20,53) nm, (66,79) nm, (19,55) nm, (66,69) nm, (51,75) nm, (59,75) nm, (85,62) nm, and (71,75), respectively. It can be seen that W and Ag give the best and worse, respectively, IMM-absorbers over this wavelength rage. The IMM-absorbers made by Cr, Ni, Ti, and W give the optimized average absorbance above 90%. The IMM-absorbers made by Ag, Al, Au, and Cu give the optimized average absorbance below 60%.

### B. Optimized IMM-absorber and its spectrum

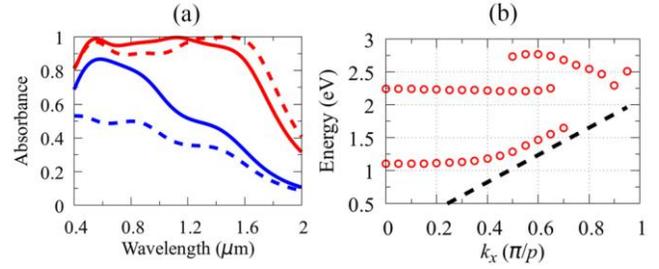

Fig. 4 (a) Absorption spectra of (i) W plane without both grating and alumina layer (blue dashed line), (ii) W plane without grating but with an alumina layer thickness 75 nm placed on top of it (blue solid line), (iii) the optimized IMM-absorber made by W (red solid line), and (iv) the optimized IMM-absorber made by W with slit height changed to 150 nm (red dashed line). (b) Dispersion of the optimized IMM-absorber made by W. The black dashed line is light line in air.

Next, we explain the physical origins of the broadband nearly-perfect absorption. The interference of non-diffracted and diffracted terms of $r_0$ can lead to the perfect absorption ($r_0$=0) when their amplitudes are the same and their phase difference is equal to ±π. The quasi-particles which are responsible for these situations are explained as follows.

The blue solid line in Fig. 4(a) shows the absorption spectrum of a W plane with an alumina layer thickness 75 nm placed on top of it (but without metal grating). The high and broad absorption peak already appears near the 553 nm (2.242 eV) resonance. This peak disappears by removing the alumina layer as shown by the blue dashed line in Fig 4(a). This indicates that this resonant peak is associated with the excitation of surface wave at the alumina-W interface. The dispersion of the optimized IMM-absorber made by W in Fig. 4(b) shows that the 2.242 eV resonance has nearly flat band (second band). A small dispersion band is typically associated with the waveguide resonance [18]. Therefore, the broad resonant peak at 553 nm corresponds to the excitation of waveguide resonance supported at the interface between the alumina layer and W plane. This resonance will be referred to as insulator-metal waveguide resonance hereafter. Another high and broad peak appears at wavelength 1,121 nm (1.106 eV) by perforating the W plane with slits height 100 nm as shown by the red solid line in Fig. 4(a). This resonant peak is sensitive to $h$, and it makes the red-shift to 1,490 nm by increasing $h$ to 150 nm as shown by the red dashed line in Fig. 4(a). This resonance also shows a little dispersion for small $k_x$ as shown in Fig. 4(b) (first band), but its dispersion is stronger approaching the insulator-metal waveguide resonance by increasing $k_x$. Therefore, the broad resonant peak at 1,121 nm corresponds to the excitation of the slit waveguide resonance. The combination of these two waveguide resonances gives the broadband absorption in our absorber. It should be noted that the third dispersion band arises from the hybridization of the insulator-metal waveguide resonance and the slit waveguide resonance at large $k_x$.

## C. Distributions of normalized magnetic fields and rate of heat generations

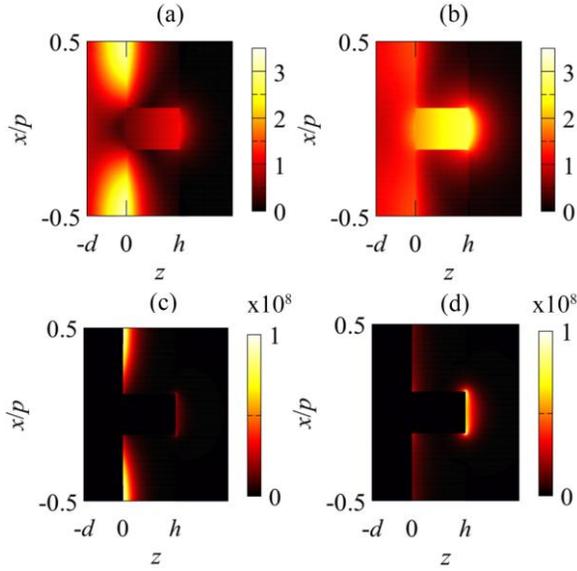

Fig. 5. (a-b) Normalized magnetic field distributions in an alumina layer, metal grating, and metal plane at wavelength 553 nm (a), and 1,121 nm (b). (c-d) Rate of heat generation normalized by incident power density at wavelength 553 nm (c), and 1,121 nm (d). All figures are obtained by using QANS with kx=0.

Figure 5(a) shows the normalized magnetic field distribution at the insulator-metal waveguide resonance of the optimized IMM-absorber made by W. The magnetic field is localized over the interface $z$=0 as expected. The magnetic field enhancement is largest at $x=\pm p/2$ which is equal to 3.2. Figure 5(b) shows the normalized magnetic field distribution at the slit waveguide resonance of the optimized IMM-absorber made by W. The magnetic field becomes localized inside the slit waveguide confirming the nature of slit waveguide resonance at this wavelength. The maximum of the magnetic field enhancement occurs at $x$=0 which is equal to 3.0. Figure 5(c)-(d) show that the rate of heat densities are strongly generated at the alumina-grating and slit-metal back plane interfaces, respectively, at the wavelengths 553 nm and 1,121 nm, respectively, which are in consistent with their corresponding magnetic field distributions. The maxima of $Q_d/S_0$ at wavelengths 553 nm and 1,121 nm are equal to $1.1 \times 10^8$ W/m$^2$ and $7.0 \times 10^8$ W/m$^2$, respectively.

## D. Average absorbance versus incident angles

Figure 6 shows the average absorbance of the optimized absorber made by W over the wavelength range 400-1,200 nm as a function of incident angles $\theta$. The averaged absorbance is above 90.0% over the incident angle range 0-66 degree. The averaged absorbance sharply drops by increasing $\theta$ beyond 66 degree. The decrease of the average absorbance results from the hybridization of the alumina-metal waveguide resonance and the slit waveguide resonance at large $k_x$. Therefore, our absorber can practically works over the relatively broad incident angle range. This angle-insensitive absorber allows wide-angle collections of incoming power from the sun light thereby increasing the absorbed energy.

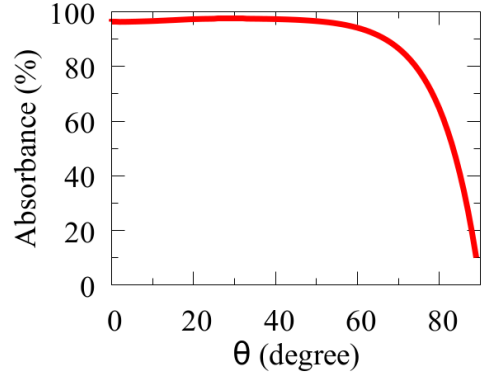

Fig. 6. Average absorbance of the optimized IMM-absorber made by W over the wavelength range 400-1,200 nm as a function of incident angles.

## 3. Discussion

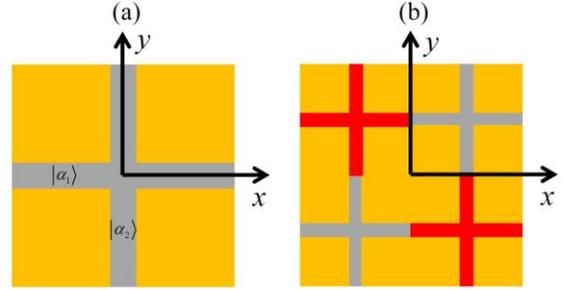

Fig.7. (a) Top view of a unit cell of two-dimensional grating made by square metal rods periodically arranged along $x$ and $y$ axes. The brown color denotes output medium as previously introduced in Fig. 1, and the gold color denotes one-fourth of square metal rod. $|\alpha_1\rangle$ and $|\alpha_2\rangle$ denote slit waveguide modes of horizontal and vertical slits, respectively. (b) Top view of a supperlattice of two dimensional grating made by smaller metal rod array. The slits indicated by the red color have smaller height than those indicated by the brown color.

Finally, we discuss the ways to improve the IMM-absorber performance. The slit array depends on the polarizations. However, its two-dimensional cousin, that is the square rod array as schematically shown in Fig. 7(a), is independent on the polarizations given the period smaller than the wavelength range of interest [19]. Since the two-dimensional grating's unit cell can be taken as two orthogonal slit waveguides $|\alpha_1\rangle$ and $|\alpha_2\rangle$ as indicated in Fig. 7(a), our results should hold for the case of the slit array replaced by the rod array. Therefore, the insulator-metal rod array-metal should provide broadband, wide-angle, and polarization independence absorber. Broader absorption extending to the infrared band can be made by using the wisdom of making more complicated unit cell [20,21]. As previously mentioned, the slit waveguide resonance makes red-shift by increasing the slit height. Therefore, we may create an absorber with broader absorption band by making multiple slits with different heights in the supperlattice as shown in Fig. 7(b). The absorber with this supperlattice should support alumina-metal waveguide resonance and multiple slit waveguide resonances thereby generating broader absorption band. Unlike the previous structure that the metallic parts of the absorber are separated, this absorber keeps all metallic parts of the absorber in touch allowing heat conduction to thermal-energy harvesting devices. The supperlattice shown in Fig. 7(b) is relatively complicated, and thus it should be simulated by using Finite Element software such as COMSOL multiphysics that supports adaptive meshes algorithm. Nevertheless, the QANS of IMM-absorber can be used to guide this complicated simulation.

## 4. Conclusion

In conclusion, we have presented the quasi-analytical solutions of wide-angle and broadband nearly-perfect IMM-absorber comprising alumina layer, metal grating, and metal back plane. The IMM-absorber made by W shows 96.2% average absorbance over the wavelength range 400-1,200 nm, and the averaged absorbance is above 90.0% for the incident angles 0-66 degrees. The broadband absorption is attributed to the excitations of two waveguide resonances at the alumina-metal grating interface and inside the slit waveguide. The absorber performance is expected to increase by replacing the metal grating with the metal rod array whose unit cell has multiple orthogonal slits with different heights.

**Funding Information.** This work is supported by King Monkut's Institute of Technology Ladkrabang [KREF145905]. C. Ruttanapun thanks the Thailand Research Fund (TRF) (Contract Number: MRG6080236) for financial support.

**Acknowledgment**. We thank the Faculty of Science, King Mongkut's Institute of Technology Ladkrabang (KMITL) for providing funding a research grant.